\documentclass{webofc}
\usepackage[varg]{txfonts}   
\usepackage{graphicx}
\usepackage{amssymb}
\usepackage{dcolumn}
\usepackage{amsmath}
\usepackage{bm}
\usepackage{textcomp}
\usepackage{epstopdf}
\usepackage{color}
\usepackage{ulem}
\usepackage[toc,page,title,titletoc,header]{appendix}
\usepackage[colorlinks,
            linkcolor=blue,
            anchorcolor=blue,
            citecolor=blue
            ]{hyperref}
\usepackage{float}
\usepackage{multirow}
\begin{document}
\title{Learning Langevin dynamics with QCD phase transition}
%
%

\author{\firstname{Lingxiao} \lastname{Wang}\inst{1} \and
        \firstname{Lijia} \lastname{Jiang}\inst{1,2} \and
        \firstname{Kai} \lastname{Zhou}\inst{1}\fnsep\thanks{\email{zhou@fias.uni-frankfurt.de}}
}

\institute{Frankfurt Institute for Advanced Studies, Ruth Moufang Strasse 1, D-60438,
Frankfurt am Main, Germany
\and
           Institute of Modern Physics, Northwest University, 710069, Xi’an, China
          }

\abstract{%
In this proceeding, the deep Convolutional Neural Networks(CNNs) are deployed to recognize the order of QCD phase transition and predict the dynamical parameters in Langevin processes. To overcome the intrinsic randomness existed in a stochastic process, we treat the final spectra as image-type inputs which preserve sufficient spatiotemporal correlations. As a practical example, we demonstrate this paradigm for the scalar condensation in QCD matter near the critical point, in which the order parameter of chiral phase transition can be characterized in a $1+1$-dimensional Langevin equation for $\sigma$ field. The well-trained CNNs accurately classify the first-order phase transition and crossover from $\sigma$ field configurations with fluctuations, in which the noise does not impair the performance of the recognition. In reconstructing the dynamics, we demonstrate it is robust to extract the damping coefficients $\eta$ from the intricate field configurations.}

\maketitle
Phase transition and critical phenomena are extensively observed in various many-body systems. The orders of phase transition or critical exponents, they can be well-extracted in equilibrium systems. However, in a non-equilibrium scenario, the situation becomes more complicated especially for a stochastic process emerges in quantum systems. The intrinsic randomness breaks the deterministic description of the dynamics, which hinders our further understanding to those exotic non-equilibrium systems, e.g., heavy-quark diffusion in the Quark-Gluon Plasma(QGP)~\cite{vanhees:2008nonperturbative,zhao:2020heavy}. The Langevin dynamics is a general description of a non-equilibrium system as a stochastic differential equation, in which the degree of freedom typically are the collective variables changing very slowly, compared to the other microscopic variables in the system. In QCD phase transitions, one challenging of both theoretical and experimental sides is if we can identify the order of the phase transition and the critical exponent of the system from the configurations generated from such a stochastic process, which is significant for the critical-point search~\cite{Mukherjee:2015swa}. Deep learning with a hierarchical structure of artificial neural networks is emerging as a novel tool to uncover structure in complex data and to efficiently describe it with a minimal number of parameters~\cite{lecun:2015deep}. 
Nowadays, the deep learning method is also utilized in the field of physics research~\cite{albertsson:2018machine,carleo:2019machine}, such as recognizing phase transitions in heavy ion collisions(HICs)~\cite{pang:2018equationofstatemeter,du:2020identifying} and improving the computation of Lattice QCD~\cite{zhou:2019regressive,blucher:2020novel,kanwar:2020equivariant}. Meanwhile, the model-free prediction on state evolution has been discussed with machine learning for chaotic dynamical systems~\cite{pathak:2018modelfree}. It makes recognizing phase transitions in stochastic processes feasible.

\begin{figure}[htbp!]
\center
\includegraphics[width=2.8 in] {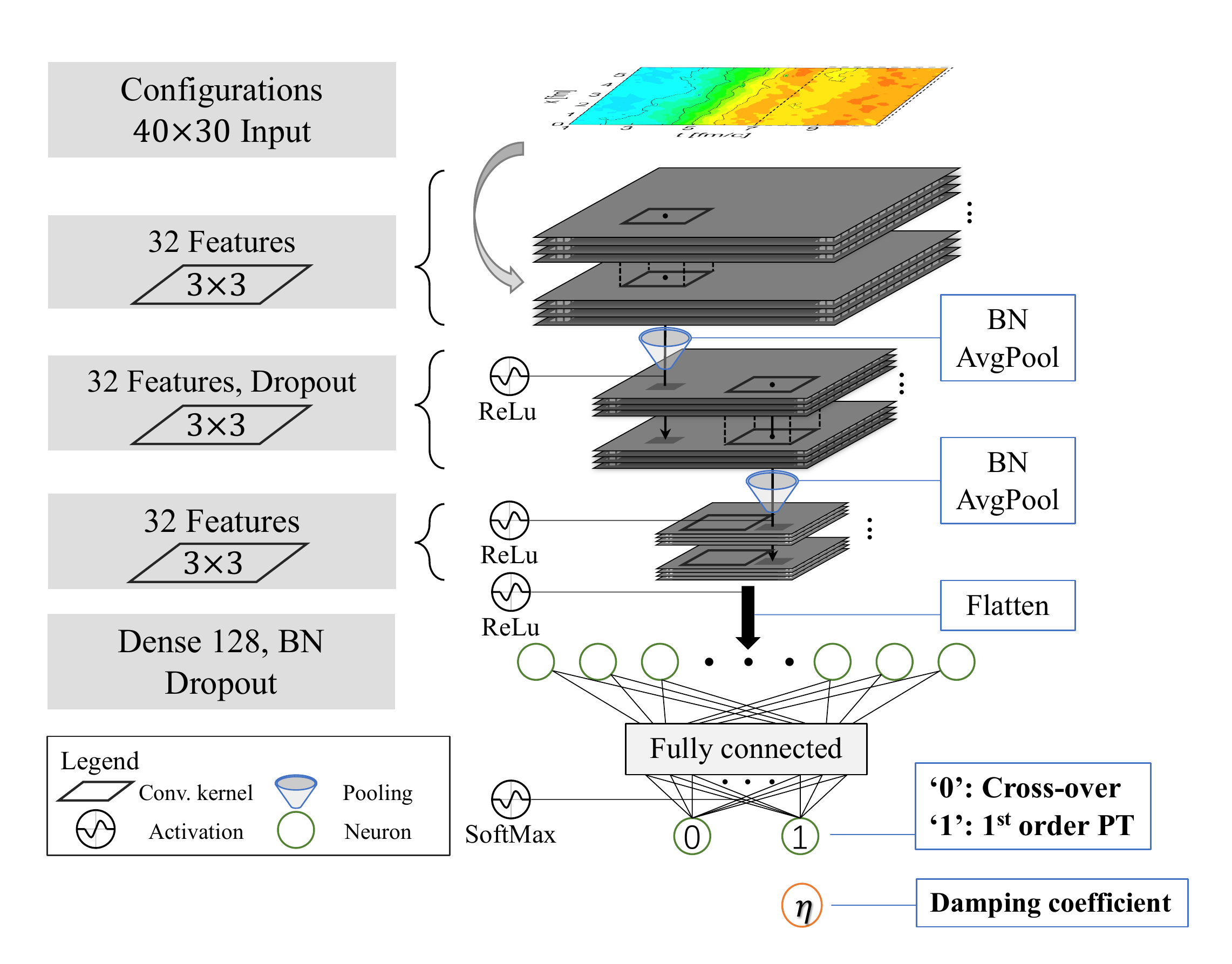}
\caption{The architecture of the deep CNNs for recognizing the phase order and predicting the damping coefficient from the $\sigma$ field configurations. Details of the network structure can be found in Ref.~\cite{jiang:2021deep}}
\label{fig:CNNs}
\end{figure}

In recent study~\cite{jiang:2021deep}, we proposed a deep CNN model to detect phase transition and dynamics in stochastic processes as Fig.~\ref{fig:CNNs} shown. In the problem set-up, the raw observations (configurations) for the system are nothing but the stochastic time series data. With regard to the effective inputs to the machine, a proper choice is the final-state particle spectra. Thus, we feed the event-by-event spatiotemporal scalar field configurations in final stage to the neural network to identify the dynamics, including the phase order and dynamical parameter as labels. As a practical example of Langevin dynamics with QCD phase transitions, we derive the dynamic equation from a linear sigma model(LSM). The effective potential from the LSM presents a scenario in which the crossover locates at the small chemical potential region and first-order phase transition occurs at the large region~\cite{roh:1998chiral}. In HICs, the hot and dense fireball created an extreme environment where the QCD phase transition can happen~\cite{bzdak:2020mapping,xu:2021littlebang}. To model the corresponding phase transition processes, the Langevin equation is adopted to describe the semi-classical evolution for the long wavelength mode of the $\sigma$ field (for more context and details can see~\cite{nahrgang:2011nonequilibrium,jiang:2017dynamical,jiang:2021deep}),
\begin{equation}
\partial ^{\mu }\partial _{\mu }\sigma \left( t,x\right) +\eta \partial
_{t}\sigma \left( t,x\right) +\frac{\delta V_{eff}\left( \sigma \right) }{%
\delta \sigma }=\xi \left( t,x\right),
\label{eq:langevin}
\end{equation}
where $\eta $ is the damping coefficient, $\xi \left( t,x\right) $ is the noise term, and the effective potential $V_{eff}$ decides the type of phase transition in the stochastic process (see more details in Ref.~\cite{paech:2003hydrodynamics}). The terms $\eta$ and $\xi$ are both from the interaction
between $\sigma$ field and the thermal background, which follow the fluctuation-dissipation theorem. In the zero-momentum mode limit, the correlation has the form $\left\langle \xi \left( t\right) \xi \left( t^{\prime }\right) \right\rangle \propto \eta \coth{\left(\frac{m_{\sigma}}{2T}\right)} \delta \left( t-t^{\prime }\right) $. In our calculation, $\eta$ is taken as a free parameter while the noise is set as white noise, and $B$ is set as the amplitude of the white noise. For simplicity but without loss of generality, we assume the heat bath evolves along trajectories with constant baryon chemical potential, and the temperature drops down in a Hubble-like way, $T(t)/{T_{0}}=({t}/{t_{0}})^{-0.45}$, where $T_{0}~(>T_c)$ is the initial temperature, and $t_{0}=1$ fm is the initial time for the evolution. With regard to the dynamical evolution of the $\sigma $ field, we set the damping
coefficient $\eta$ to be constant across the evolution with values ranging from $1.0$ to $5.5\,\text{fm}^{-1}$. As for the details of the numerical set-up, we simulate the evolution of $\sigma$ field in $1$-dimensional space with range $L=6.0$ fm, and the spatial grid size $dx = 0.2$ fm. The duration of the evolution is $16$ fm/c at most with the temporal step size $dt=0.1$ fm/c. With the above set-ups, the $\sigma$ field was evolved according to Eq.~\eqref{eq:langevin} on an event-by-event basis. The configurations from later episode after phase transition with $4$ fm duration of the $\sigma$ field are censored as the input data-sets. Thus, the input is in $t\in[7, 11]\,\text{fm}^{-1}$, or in the last 40 time-steps from the evolution, where the ambient temperature is already much lower than $T_c$, ensured that the potential phase transition already happened. Therefore, the prepared input configuration contains $N= 40\times 30 = 1200$ pixels in each event. 

\begin{figure}[htbp!]
\center
\includegraphics[width=5 in]{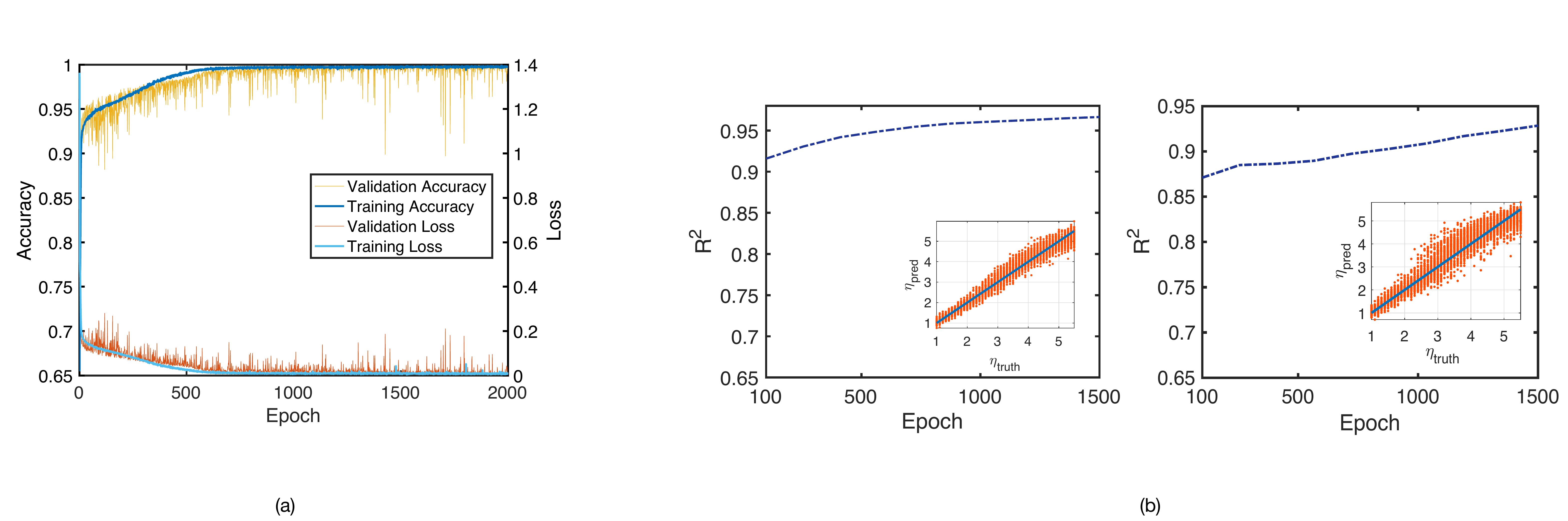}
\caption{(a) The accuracy and loss on training and validation data sets with different noise parameters $B$=0.5, 1, 1.5, 2 and 2.5. (b) The training process in predicting the damping coefficient from configurations. The left panel are prediction results for the crossover scenario, the right panel are prediction results for the first-order scenario.}
\label{fig:acc_train}
\end{figure}
To train the neural networks, we prepare 10 000 events at each parameter setup with labels. First we deploy the machine on data-sets with B = 0.5 and 1, and the training history is recorded in Fig.\ref{fig:acc_train}(a). The validation loss tends to decrease same as training loss with epoch increasing, in other words, it reaches stable without over-fitting. Moreover, the testing accuracy reaches to 99.9\%, it means the performance of the well-trained machine is robust and these final-stage configurations indeed carry with phase transition information. It should be noticed that, although the configurations are totally different from the learned ones due to the intense noise, the machine can keep an accurate prediction on the phase order in each event. As for the damping coefficient, as the core dynamic parameter, it can influence not only the thermal diffusion speed of the system but also the fluctuations. Here we use events generated on both the crossover and the first-order scenarios with the damping coefficient in the range of $(1.0-2.5)$ and $(4.6-5.5)\,\text{fm}^{-1}$, in each $\eta$-bin, 1000 events were collected as inputs. In the training process of Fig.~\ref{fig:acc_train}(b), the related coefficient $R^2$ grows quickly from 87\% to 93\% for the crossover and from 91\% to 97\% for the first-order scenario. The evaluation on the testing data-set is shown in the insert part. It consists of orange dots which are labeled by the damping coefficients of the ground truth and predictions on testing data-set from the trained CNN. If we take a close step to see the performance of the machine. It is found that although inside the training set has no supervision in the region $(2.6 - 4.5)\,\text{fm}^{-1}$, the predictions still keep a reasonable performance. 

In summary, we demonstrate the framework is effective in recognizing the QCD phase transitions and extracting the Langevin dynamics from complicated configurations. The first-order or crossover-type phase transitions, encoded inside the stochastic dynamical evolution, can be detected by a well-trained machine. Although the field configurations differ totally from each other because of the noise terms and also the random initial conditions, the machine successfully learns to make an accurate prediction on the phase order for previously unseen evolution events in the testing stage. This is related to the powerful capability of the deep CNNs for extracting hidden correlations in image-type data-set, which facilitates the presented phase order identification from the field configurations.We further prepare mixed configurations with damping coefficient in the range of $(1.0-2.5)\text{fm}^{-1}$ and $(4.6-5.5)\text{fm}^{-1}$, nevertheless, the predictions to the dynamics are made for the entire range. It reveals an acceptable generalization ability when tested on configurations containing damping beyond the training set. The present method can be helpful for a broader field, like, there is a potential application in topological-dependent stochastic process~\cite{zhao:2020topologydependent}, in which the topological charge could be extracted by the deep CNNs. A recent attempt is to detect the chiral magnetic effects in HICs with the same paradigm~\cite{zhao:2021detecting}.

\section*{Acknowledgment}
The authors acknowledge inspiring discussions with Horst Stoecker.
The work is supported by the AI grant at FIAS of SAMSON AG, Frankfurt (L. J., L. W. and K. Z.), by the BMBF under the ErUM-Data project (K. Z.),  and by the NVIDIA Corporation with the generous donation of NVIDIA GPU cards for the research (K. Z.).

\bibliography{langevin}

\end{document}